\documentclass[12pt]{article}
\usepackage{graphicx}
\begin{document}

\title{Phase transitions in isolated strongly interacting systems}

\author{Martin Veselsky\\
\\
Institute of Physics, Slovak Academy of Sciences,\\
Dubravska cesta 9, Bratislava, Slovakia\\
e-mail: fyzimarv@savba.sk
}

\date{}

\maketitle

\begin{abstract}

Thermodynamical properties of nuclear matter at sub-satura\-tion densities 
were investigated using a simple van der Waals-like equation of state 
with an additional term representing the symmetry energy. 
First-order isospin-asymmetric liquid-gas 
phase transition appears restricted to isolated isospin-asymmetric 
systems 
while the symmetric systems will undergo fragmentation 
decay resembling the second-order phase transition. 
The density dependence of the symmetry 
energy scaling with the Fermi energy 
satisfactorily describes the symmetry energy at sub-saturation 
nuclear densities. 
The deconfine\-ment-confine\-ment phase transition from the quark-gluon 
plasma to the confined quark matter appears in the isolated systems 
continuous in energy density while discontinuous in quark 
density. A transitional state of the confined quark matter has 
a negative pressure and after hadronization an explosion scenario 
can take place which can offer explanation for 
the HBT puzzle as a signature of the phase transition.  
\end{abstract}

\section*{Introduction}

\indent 

The knowledge of the phase diagram of nuclear matter is one of the 
principal open questions in modern nuclear physics 
with far reaching cosmological consequences. Detailed 
investigations have been carried out in the recent years in particle-nucleus 
and nucleus-nucleus collisions in a wide range of projectile energies.   

The process of multifragmentation was investigated at intermediate and 
high energies in order to study the properties of the expected 
liquid-gas phase transition at sub-saturation nuclear densities. 
For instance, using the calorimetry of the hot quasi-projectile nuclei formed 
in the damped nucleus-nucleus collisions (see Ref. \cite{MVCorrSig} 
and Ref. \cite{MVIsoTrnd} for a review of related methods), 
thermodynamical observables of the fragmenting system 
such as temperature and chemical potential were extracted 
and a set of correlated signals of the isospin-asymmetric 
liquid-gas phase transition was observed.  
On the other hand, various properties of the observed fragments 
can be interpreted as signals of criticality 
\cite{Elliot,Zipf,NimCrit}, thus indicating an 
underlying second order phase transition. The phase diagram at sub-saturation 
densities appears to be rather complex and detailed investigations 
are needed for clarification. 

The recent experimental 
studies of the nucleus-nucleus collisions at ultrarelativistic energies 
provide vast amount of experimental observations which can be interpreted 
as signals of production and decomposition of the new state of matter 
with deconfined quarks, called quark-gluon plasma (QGP) \cite{HeinzRev}. 
Such matter exhibits many unique properties 
such as nearly perfect fluid behavior \cite{Fluid} 
and strong medium modification 
of high-energy particles with a typical Mach cone behavior \cite{MachCone}. 
One of the phenomena not yet understood there is the so-called 
"HBT puzzle" \cite{HBT}, 
an unexpected behavior of the source radii extracted from 
particle-particle correlations, which appear to be almost isotropic, 
in contrast with intense elliptic flow observed in such 
collisions. It was suggested that spatial inhomogeneity 
(typically in the form of QGP droplets) can play a crucial role in its 
explanation \cite{QCDDrop}. The mechanism of formation 
of such inhomogeneity is one of the interesting open questions.

In the present work, the mechanism of a phase transition is 
investigated as a phenomenon reflecting the properties of the 
underlying interaction. A modified van der Waals equation of state 
is used for the nuclear medium at sub-saturation densities and 
the role of isospin asymmetry will be investigated within the limitation 
following from the evolution of an isolated system.  
An analogous treatment is used for schematic description 
of the system undergoing deconfinement-confinement 
phase transition. 

\begin{figure}[h]                                        

\centering
\includegraphics[width=12.0cm, height=6.0cm ]{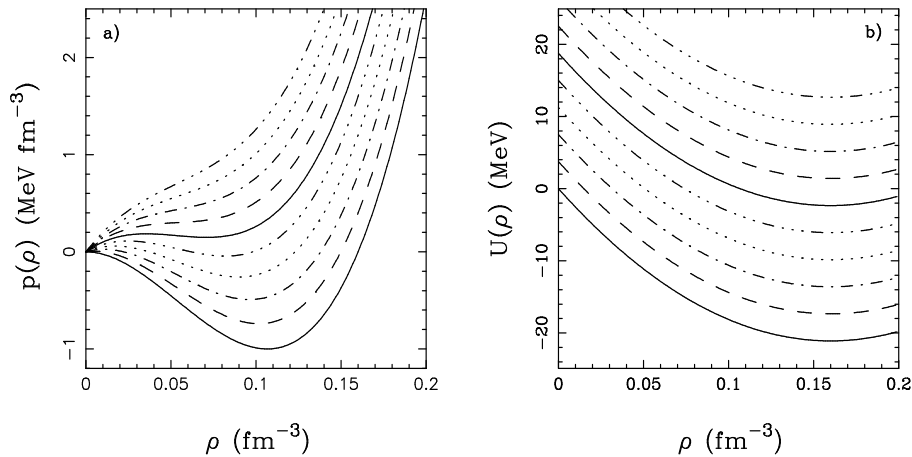}

\caption{
Density dependence of a) pressure and b) internal energy, 
given by the nuclear equation of state from Eq. \ref{NucEoS}.  
The lines represent isotherms 
with temperatures increasing with the step 2.5 MeV, 
starting at zero.
           }
\label{fg1}
\end{figure}

\section*{Isospin asymmetric liquid-gas phase transition in the nuclear matter}

The liquid-gas phase transition in the nuclear matter is supposed to occur 
after nuclear matter expands and medium becomes unstable toward 
density fluctuations. The process as initially suggested in Ref. \cite{Sauer} 
was supposed to be analogous to liquid-gas phase transitions in real 
gases, which is typically described by the van der Waals equation of state 
(EoS). However, unlike the van der Waals gas, nuclear matter 
is a two-component 
system and it thus possesses an additional degree of freedom related to 
proton and neutron concentrations. Additional energy term representing 
the symmetry energy has to be taken into account. Under assumption 
that the symmetry energy decreases with density, it 
can be favorable for the expanding system to store the excess of neutrons 
(protons) 
into a dilute phase and thus isospin-asymmetric liquid-gas phase transition 
can occur \cite{Muller}. However, the expanding nuclear 
system is isolated 
and no heat can be transferred into or out of the system and a 
typical picture of phase transition under conditions of phase equilibrium, 
as represented by the well-known Maxwell construction, is not relevant due to 
necessity to transfer latent heat into or out of the system.  

In order to study the mechanism of isospin asymmetric liquid-gas phase 
transition in the isolated nuclear system, we adopt a simple 
EoS for the symmetric nuclear matter, closely reminding the 
van der Waals EoS, where the repulsive 
short-range cubic interaction term is introduced as a substitute 
for the minimum volume 

\begin{equation}
p = \rho T + a\rho^2 + b\rho^3
\label{NucEoS}
\end{equation}

The parameter values $a = - 263.7 \hbox{ }\rm{MeV fm^3}$ i
and $b = 1674.94 \hbox{ }\rm{MeV fm^6}$
were chosen to fix zero pressure at saturation density 
taken as $\rho_0 = 0.16 \hbox{ }\rm{fm^{-3}}$ and preserve plausible values 
for both critical temperature ($T_c \simeq 14 \hbox{ }\rm{MeV}$) 
and ground state 
energy ($U_{gs} \simeq -21 \hbox{ }\rm{MeV}$). 
The equation of state thus can be considered 
as a reasonable approximation to the results obtained using the 
microscopic theory of nuclear matter.  
The resulting pressure and internal energy isotherms 
are shown in Fig. \ref{fg1}. The internal energy 
(obtained assuming the heat capacity $c_V = 3/2$) exhibits a quadratic 
dependence on density with a minimum at saturation density, what is the 
expected result for the nuclear ground state.

\begin{figure}[htbp]                                        

\centering
\includegraphics[width=12.0cm, height=10.0cm ]{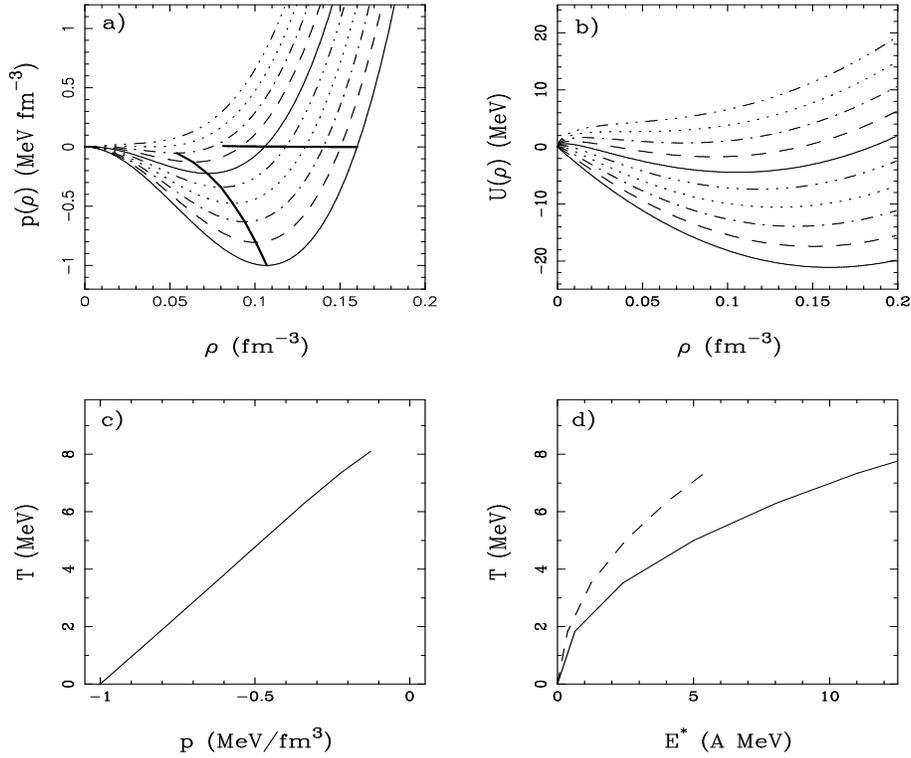}

\caption{
Density dependence of a) pressure and b) internal energy 
for constant entropy. 
The lines represent adiabatically expanding systems  
with the temperature at saturation density $T_0$ increasing with 
the step 2.5 MeV, starting at zero.
Curve in the panel a) represents the spinodal contour and the 
horizontal line heating process of the nucleus.  
In the panel c) is the p-T diagram for the spinodal 
contour and in the panel d) are the caloric curves for the 
spinodal contour (solid) and for the Fermi gas formula (dashed). 
           }
\label{fg2}
\end{figure}

Since the system is thermally isolated, the expansion of the system 
is an adiabatic process. The entropy does not change during 
system evolution and assuming that the heat capacity $c_V$ is constant, 
the following expression for temperature as a function of density 

\begin{equation}
T = T_0 \left(\frac{V_0}{V}\right)^{1/c_{v}} 
= T_0 \left(\frac{\rho}{\rho_0}\right)^{1/c_{v}}
\label{TAdiab}
\end{equation}

can be derived, 
where $T_0$ is the temperature projected to saturation density $\rho_0$. 
The resulting isoentropic density dependences of the pressure and internal 
energy 
are shown in Fig. \ref{fg2}a, b. One can identify spinodal region 
where pressure grows while density falls, however, the limiting temperature 
appears much smaller than the one extracted from Fig. \ref{fg1}. 
The resulting phase diagram 
constructed using the pressure and temperature at the spinodal 
contour is shown in Fig. \ref{fg2}c. The critical temperature appears 
to only slightly exceed 8 MeV. Such observation is in agreement 
with experimental results \cite{JBNLimTmp}.  
Two caloric curves are shown in Fig. \ref{fg2}d. 
The dashed line represents the standard caloric curve obtained 
using the Fermi-gas formula 

\begin{equation}
E^* = \tilde{a} T^2 
\hbox{ ,}
\label{fgf}
\end{equation}

where the level density 
parameter is chosen as $\tilde{a} = A/(10 \hbox{ }\rm{MeV})$. 
Such dependence 
is widely used in various evaporation models and represents essentially 
an evolution at ground state density. It is worthwhile to note 
that the temperature dependence in Eq. \ref{fgf} is equivalent to the 
heat capacity $c_V$ linearly increasing with temperature. Such 
dependence is in agreement with theory of the Fermi gas at low excitation 
energies, however, its validity should be restricted only to the 
domain where $c_V$ does not exceed the ideal gas value 3/2 (below 
temperature 7.5 MeV), since  
$c_V$ saturates at this value and thus assumption of its further 
growth is not correct.  
Solid line in Fig. \ref{fg2}d 
represents the caloric curve obtained by combining the temperature 
at which the system enters 
spinodal region with the estimate of initial excitation energy 
taking into account also expansion of the system during initial heating. 
The heat capacity $c_p$ was obtained from 

\begin{equation}
c_p = c_V  +  \frac{T}
{ T + 2a\rho + 3b\rho^2 }
\hbox{ ,}
\label{cp}
\end{equation}

where $c_V$ was determined using the formula \ref{fgf}. 
The system expanded according to 

\begin{equation}
\frac{1}{V}\left(\frac{\partial V}{\partial T}\right)_p 
= \frac{1}{ T + 2a\rho + 3b\rho^2 }
\hbox{ ,}
\label{expcoeff}
\end{equation}

which is closely related to $c_p$. At each step of the procedure 
the temperature was incremented by a small value, corresponding increase 
of the internal (excitation) energy was proportional to $c_p$ and 
the density was modified according to the calculated volume expansion 
coefficient. Finally, new value of the pressure was calculated from  
the equation of state using new values of density and temperature. 
The resulting evolution of the system is shown in Fig. \ref{fg2}a 
as a nearly horizontal line starting from the ground state. The 
pressure of the system appears to grow very slowly while the system 
expands. The resulting caloric curve is similar but much flatter 
than the one obtained using the Fermi gas formula. For real 
nuclei one can assume that the system will follow the dashed 
line until it has enough energy to pass the fragmentation barrier 
and expand toward spinodal region, thus approaching the solid 
line via a short plateau. Such a short plateau was observed in 
the experiment \cite{MVCorrSig}, where it correlated with 
the onset of thermal equilibrium with the nucleonic gas, as represented 
by agreement of the extracted double isotope-ratio temperature with the 
kinematic temperature of protons.  
   
Furthermore, it is interesting to note that while the fermionic system 
expands its Fermi energy exhibits the $\rho^{2/3}$ density dependence, 
which is identical to density dependence of temperature during adiabatic 
expansion when one assumes the saturation value $c_V = 3/2$. Thus the ratio 
of the Fermi energy to temperature, which is a parameter determining 
the occupation numbers, will remain the same during adiabatic expansion. 
The ideal gas approximation $c_V = 3/2$ appears justified for the 
isoentropic expansion where such value was reached during heating, 
in the Fig. \ref{fg2}a such situation applies to pressure curves 
with initial temperatures above $T_0 = 9 \hbox{ }\rm{MeV}$. 

Nuclear matter is a two-component system and thus 
an energy term representing 
the symmetry energy has to be introduced in order to reflect 
the difference 
of proton and neutron concentrations 

\begin{equation}
\frac{E_{asym}}{A} 
=  a_a (\frac{\rho}{\rho_0})^{2/3} \left(\frac{N-Z}{A}\right)^2 
\hbox{ ,}
\end{equation}

where $a_a$ is the symmetry energy coefficient from the 
Weisz\"acker formula and a $\rho^{2/3}$ density dependence 
is assumed from the behavior of the Fermi energy. 
Due to density dependence of the symmetry energy it  
can be favorable for the system to store the excess of neutrons ( protons ) 
into a dilute phase and thus the isospin-asymmetric liquid-gas phase transition 
can take place. The equation of state will be modified to 
the form 

\begin{equation}
p = \rho T + a\rho^2 + b\rho^3
+ \frac{2a_a \rho_0}{3}(\frac{\rho}{\rho_0})^{5/3}\left(\frac{N-Z}{A}\right)^2
\hbox{ ,}
\label{AsNucEoS}
\end{equation}

which allows us to investigate the behavior of the system as a function of 
density and isospin asymmetry $I = (N-Z)/A$. The additional term 
in the equation of state will not modify the temperature dependence 
in Eq. \ref{TAdiab}. 

\begin{figure}[h]                                        

\centering
\includegraphics[width=12.0cm, height=6.0cm ]{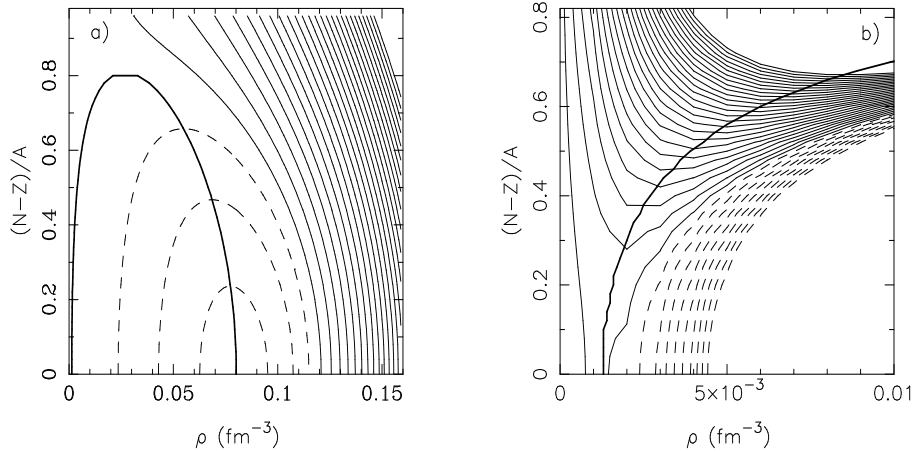}

\caption{
a) Pressure contour plot, corresponding to the nuclear 
equation of state given in Eq. \ref{AsNucEoS}.  
The adiabatic expansion  
with the temperature at saturation density $T_0 = 10 MeV$ is assumed.
Thick curve represents the spinodal contour. In panel b) is shown a 
detailed view onto low density region. 
           }
\label{fg3}
\end{figure}

The pressure contours determined by Eq. \ref{AsNucEoS} 
using the value $a_a = 25 \hbox{ }\rm{MeV}$ are shown in Fig. 
\ref{fg3} as a contour plot. The resulting 
spinodal region is highlighted and its shape is similar 
to microscopic calculations \cite{SpinCont}. 
The asymmetric systems will 
enter the spinodal region at lower densities, on the other hand 
the isospin-asymmetric gas remains stable at much higher densities 
than the symmetric one. 

\begin{figure}[h]                                        

\centering
\includegraphics[width=12.0cm, height=6.0cm ]{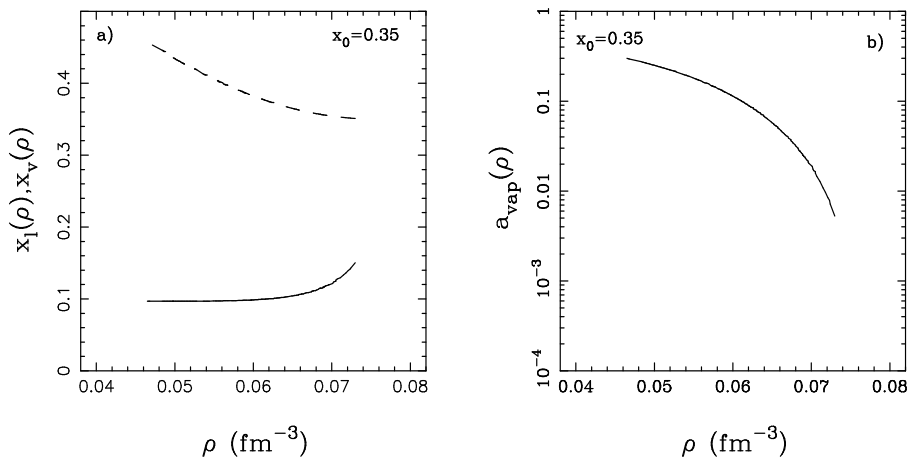}

\caption{
The resulting values of a) proton concentrations $x_v$ (solid line), 
$x_l$ (dashed line) and 
b) vapor fraction $a_{vap}$ for 
an adiabatically expanding system with $T_0 = 10 \hbox{ }\rm{MeV}$ 
and isospin asymmetry 
of the system $(N-Z)/A = 0.3$.  
           }
\label{fg4}
\end{figure}

Metastable state inside the spinodal contour 
($( \partial p / \partial V )_S > 0$)
is unstable toward spontaneous compression to stable liquid state at the
spinodal contour, but such process would create
heat. Instead of heat, the released internal energy can be
deposited into expansion of vapor into free volume, thus spending
released energy into overcoming attractive nuclear force.
When the whole system is isolated ($\delta Q_{tot} = 0$) 
the process of spinodal decomposition will be equivalent 
to fast transfer of particles from the metastable state at its 
constant pressure to the stable liquid and vapor 
states at their corresponding pressures and total 
enthalpy of the system will not change during phase transition 
and the relation 

\begin{equation}
U_{ms} + p_{ms} V_{tot} = U_{liq} + p_{liq} V_{liq} + U_{vap} + p_{vap} V_{vap}
\end{equation}

will be preserved. 

\begin{figure}[h]                                        

\centering
\includegraphics[width=12.0cm, height=6.0cm ]{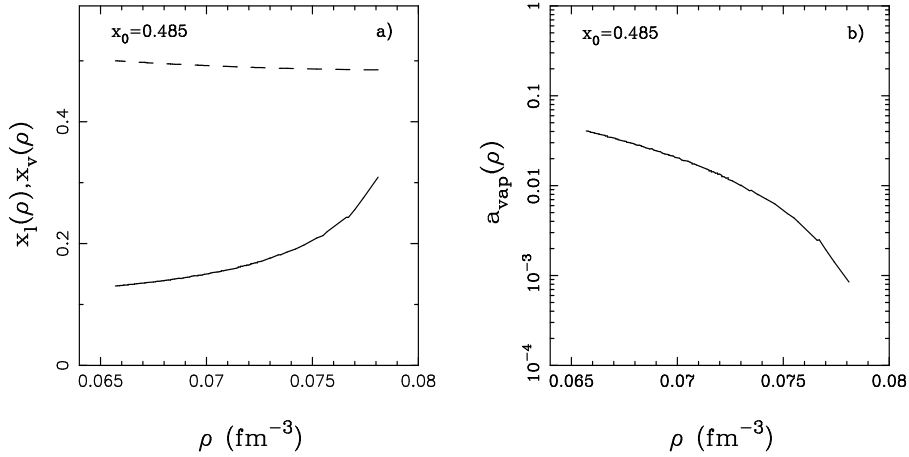}

\caption{
The resulting values of a) proton concentrations $x_v$ (solid line), 
$x_l$ (dashed line) and 
b) vapor fraction $a_{vap}$ for 
an adiabatically expanding system with $T_0 = 10 \hbox{ }\rm{MeV}$ 
and isospin asymmetry 
of the system $(N-Z)/A = 0.03$.  
           }
\label{fg5}
\end{figure}

During fast transition the total volume remains constant ($dV_{tot} = 0$)
and for a given combination of the stable liquid and gas states located
at the spinodal contour the partial volumes of liquid and vapor are

\begin{equation}
V_{tot} = \frac{A_{tot}}{\rho_{ms}} = V_{liq} + V_{vap} = \frac{A_{liq}}{\rho_{liq}}
+ \frac{A_{vap}}{\rho_{vap}}
\hbox{ ,}
\end{equation}

where

\begin{equation}
A_{liq} = A_{tot} - A_{vap}
\end{equation}

and nucleonic fraction of the vapor is

\begin{equation}
a_{vap} = \frac{A_{vap}}{A_{tot}}
= \frac{\rho_{vap}}{\rho_{ms}}\frac{\rho_{liq}-\rho_{ms}}
{\rho_{liq}-\rho_{vap}}
\end{equation}

for which the charge conserving proton concentration
($x = Z / A$) is
determined using proton concentration of a given liquid state $x_l$

\begin{equation}
x_{vc} = \frac{x_{ms}-(1-a_{vap})x_l}{a_{vap}}
\hbox{ ,}
\end{equation}

which should agree with the proton concentration $x_v$ 
of the vapor state at the spinodal contour used to determine $a_{vap}$.
Such a self-consistent stable liquid and vapor states
located at the spinodal contour can be determined numerically.
Total enthalpy conservation relation 
can be written as 

\begin{equation}
H_{ms} - H_l - H_v = h_{ms} - a_{vap} h_{v} - (1 - a_{vap}) h_{l} = 0
\hbox{ ,}
\end{equation}

where $h_{ms}$, $h_l$ and $h_v$ are values of the enthalpy per particle.

\begin{figure}[h]                                        

\centering
\includegraphics[width=12.0cm, height=6.0cm ]{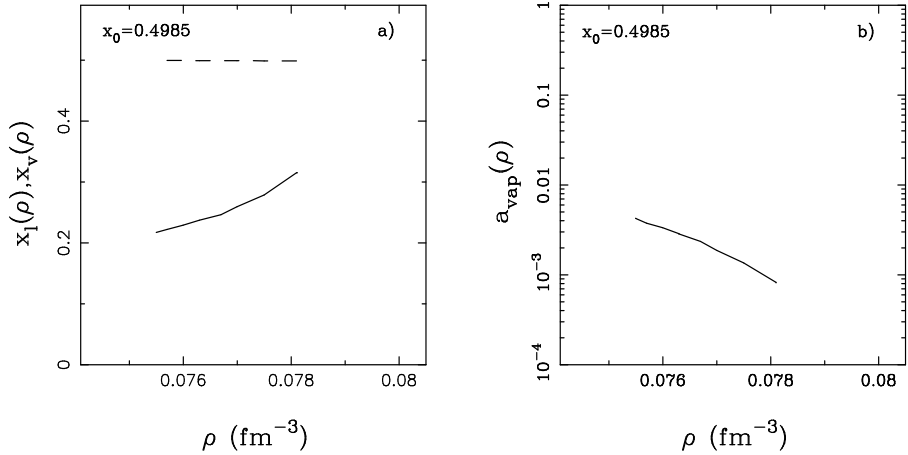}

\caption{
The resulting values of a) proton concentrations $x_v$ (solid line), 
$x_l$ (dashed line) and 
b) vapor fraction $a_{vap}$ for 
an adiabatically expanding system with $T_0 = 10 \hbox{ }\rm{MeV}$ 
and isospin asymmetry 
of the system $ (N-Z) / A = 0.003$.  
           }
\label{fg6}
\end{figure}

The resulting values of $x_v$, $x_l$ and $a_{vap}$ for 
an adiabatically expanding system with $T_0 = 10 \hbox{ }\rm{MeV}$ 
are shown in Figs. 
\ref{fg4}, \ref{fg5} and \ref{fg6} for the isospin asymmetries 
of the system $(N-Z)/A =$ 
0.3, 0.03 and 0.003, respectively. It can be seen that 
for the asymmetric system with $(N-Z)/A =$ 0.3 
there occurs a well pronounced isospin 
asymmetric liquid-gas phase transition with a sizable part of the 
system being converted into very isospin asymmetric vapor while 
the rest of the system becomes still more and more symmetric liquid. 
These trends become more prominent as the density of initial meta-stable 
state decreases. 
When the system becomes more symmetric (see Figs. \ref{fg5} and \ref{fg6}) 
the vapor fraction drops very quickly while the vapor preserves its 
isospin asymmetry. Also the region of densities where isospin
asymmetric liquid-gas phase transition can occur shrinks quickly. 
From the Figs. \ref{fg4}, \ref{fg5} and \ref{fg6} one can conclude 
that observable isospin
asymmetric liquid-gas phase transition will occur for very asymmetric systems 
such as these observed in the experiment \cite{MVCorrSig} 
while for symmetric systems (and for the remnants of liquid phase 
in asymmetric systems) another mechanism will play role. In order to reach 
spinodal region the system must pass the multifragmentation barrier and thus 
these systems will be unstable toward sudden decomposition into multiple 
fragments. Such a scenario can provide 
explanation for the somewhat confusing situation where signatures 
of both first and second order phase transitions 
can be extracted from observed  data (even from the same 
set of data). The first-order behavior can be attributed to the 
distillation of isospin-asymmetric gas while the second-order 
signatures can be attributed to subsequent fragmentation (percolation) 
of the remaining symmetric liquid. 

It is also interesting that 
the conclusions observed here are in good agreement with the assumptions 
used by the Statistical Model of Multifragmentation (SMM) \cite{SMM} 
where the fragment partition is considered at equilibrium with 
nucleonic gas. Such a scenario is supported also by experimental data 
\cite{MVCorrSig}. Other interesting point is that the $\rho^{2/3}$ density 
dependence of the symmetry energy used in the present work appears 
as satisfactorily describing the properties of asymmetric nuclear matter 
at sub-saturation densities.

\section*{Deconfinement-confinement phase transition in the quark matter}

The ultrarelativistic nucleus-nucleus collisions result in the 
very high energy densities where the quark matter can exist 
in thermally equilibrated deconfined state known as quark-gluon 
plasma. During the subsequent expansion energy density decreases 
until the critical value around 1 $GeV fm^{-3}$ where, according 
to the predictions of the quantum chromodynamical calculations 
performed on the lattice \cite{QCDLatt} 
the matter will revert to the confined 
state, where the interaction is dominated by a 
divergent confinement potential, linearly increasing 
with the distance. The potential energy, stored in the 
one-dimensional field tubes, called strings, 
will be subsequently released by string fragmentation 
in the form of hadronic gas. In order to investigate the phase 
diagram, one can use known equations of state of initial and final states 
of the quark-gluonic plasma and the hadronic gas and try to interpolate 
the equation of state in the spinodal region, as is done 
in Ref. \cite{RandQGP}. However, since the process of hadronization 
typically proceeds via fragmentation of elongated strings 
it is also interesting to explore 
thermodynamical properties of the possible intermediate state of matter, 
where the particles of the quark gas interact via confinement potential. 
Such state of matter is referred here as the confined quark matter. 
The properties of intermediate states can be important 
for the mechanism of phase 
transition, as was demonstrated in the case of liquid-gas 
phase transition at sub-saturation densities. Here we will 
explore thermodynamical properties of the confined quark matter 
in analogous way as thermodynamical properties of the asymmetric nuclear 
matter in previous section.    

In order to investigate properties of the confined quark matter 
a confinement potential is defined as 

\begin{equation}
V_{conf}(r) = \kappa r
\hbox{ ,}
\end{equation}

where $\kappa$ is the string tension with a typical value 
of 1 GeV/fm, as used in the string fragmentation models \cite{Lund}. 
In order to explore properties of the confined quark matter, a 
3-dimensional cubic lattice with a quark-quark distance $L$ 
can be assumed, where each quark interacts with six nearest  
sites. Then the component of the internal energy, corresponding 
to confinement potential, can be determined as 
 
\begin{equation}
U_{conf} = \frac{1}{2} N_q (6 \kappa L) = 3 N_q \kappa \rho^{-1/3}
\hbox{ ,}
\end{equation}

where the number of quarks $N_q$ is large enough to neglect  
surface effects. 
However, an assumption that quark on the lattice interacts 
with all neighbors must not necessarily be satisfied, since 
strings are essentially one-dimensional objects. One can 
modify above formula by assuming 
that each quark on the lattice interacts with  
$w$ neighboring quarks and one obtains the formula 
 
\begin{equation}
U_{conf} = \frac{1}{2} N_q (w \kappa L) 
= N_q \kappa \frac{w}{2}  \rho^{-1/3}
\hbox{ .}
\label{uconf}
\end{equation}

The pressure component corresponding to confinement potential 
can be obtained as 

\begin{equation}
\Delta p_{conf} = - \frac{\partial U_{conf}}{\partial V} = 
- \kappa \frac{w}{6} \rho^{2/3} 
= - \frac{1}{3} \frac{U_{conf}}{V}
\hbox{ ,}
\label{pconf}
\end{equation}

thus leading to negative pressure. In order 
to obtain the equation of state 
of the confined quark matter, one has to introduce 
the pressure of the quark gas which in the ultrarelativistic 
limit can be written as 

\begin{equation}
p
= \frac{\epsilon}{3}
= \frac{8 \pi g_q}{h^3}
T^4 \eta(4) 
= \frac{8 \pi g_q}{h^3}
T^4 \frac{7 \pi^4}{720}
\hbox{ ,}
\label{purgs}
\end{equation}

where $g_q$ is a degeneration factor and $\eta$ is the Dirichlet 
Eta function. In order to write 
equation of state with the confinement potential one needs to 
find relation between energy density and quark density. 
That is possible for zero chemical 
potential (with $g_q$ accounting for equal numbers of quarks 
and anti-quarks)

\begin{equation}
\rho = \frac{N}{V} 
= \frac{8 \pi g_q}{h^3}
T^3 \eta(3)
\label{durgs}
\end{equation}

and after comparing the equations (\ref{purgs}) and (\ref{durgs}) 
one arrives to

\begin{equation}
p = \rho T \frac{\eta(4)}{\eta(3)} 
- \kappa \frac{w}{6} \rho^{2/3} 
\end{equation}

for the equation of state. 

\begin{figure}[h]                                        

\centering
\includegraphics[width=12.0cm, height=6.0cm ]{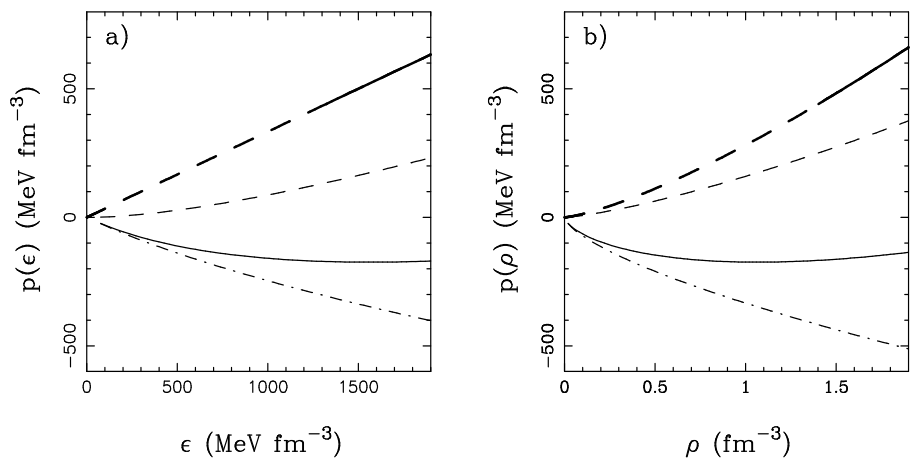}

\caption{
Dependence of the pressure on a) energy density and b) quark density 
for $w = 2$. 
Thick line represents quark-gluon plasma 
(Eqs. (\ref{purgs}) and (\ref{durgs}) with $g_q = 24$ and $g_g = 16$), 
solid section corresponds to the region above phase transition. 
Thin solid line shows pressure of the confined quark matter, 
dashed line shows the quark gas contribution  
(Eqs. (\ref{purgs}) and (\ref{durgs}) with $g_q = 24$), 
dash-dotted line shows the confinement potential 
contribution (Eq. (\ref{pconf})). 
           }
\label{fg7}
\end{figure}

The resulting situation is shown in Fig. \ref{fg7} for $w = 2$. 
At energy densities below critical value 1 $GeV/fm^{3}$ 
and corresponding quark densities the confined quark matter 
exhibits negative pressure and also violates the mechanical 
stability criterion. Such a medium is indeed unstable against 
collapse what on the other hand is a direct consequence of the 
interaction rather than an interplay 
of short- and long-range interactions, as in the 
liquid-gas phase transition. Since confined quark matter is primarily 
unstable toward hadronization it can be assumed that the survival 
time against hadronization is much shorter than collapse time 
and the medium can be considered as a quasi-stable transitional state. 
 
The detailed mechanism of the deconfinement-confinement phase transition 
is beyond the scope of this work, the lattice QCD calculations \cite{QCDLatt} 
suggest a rapid crossover for the systems with small chemical potential 
produced at RHIC.  
As in the liquid-gas phase transition at sub-saturation densities, 
one can explore the global properties of the interaction and 
limitations on the phase transition 
resulting from the isolated system and, as a consequence, 
an isoenthalpic transition into confined quark matter can be considered. 
In the present case the hot quark-gluon plasma 
with high thermal pressure transforms into confined quark matter 
with much lower (negative) pressure. The hadronization 
phase transition then follows as a consequence of the 
fragmentation of the elongated strings. 

\begin{figure}[h]                                        

\centering
\includegraphics[width=12.0cm, height=10.0cm ]{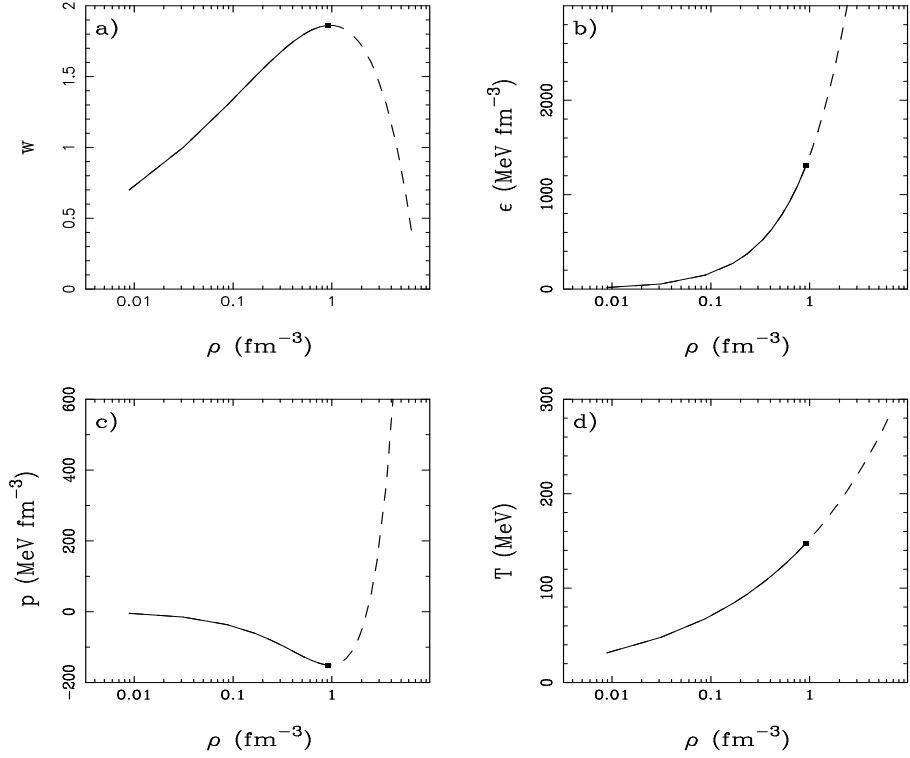}

\caption{
Solutions of the Eq. \ref{isenthnum}. 
Dependence of the a) parameter $w$, 
b) energy density, c) pressure and  
d) temperature of the confined quark matter on quark density. 
Solid lines  represent the low-density branch, 
dashed lines show the high-density branch.  
           }
\label{fg8}
\end{figure}

Enthalpy of the quark-gluon plasma with $u$ and $d$ quarks 
and antiquarks can be expressed as 

\begin{equation}
U + pV = 4 N_q ( \frac{\eta(4)}{\eta(3)} + 
\frac{g_g \zeta(3)}{g_q \eta(3)} \frac{\zeta(4)}{\zeta(3)} ) T
\hbox{ ,}
\end{equation}

where $g_q$ = 24, $g_g$ = 16 and $\zeta$ 
is the Riemann Zeta function. 
The enthalpy balance of the transition from quark-gluon plasma at temperature 
$T_0$ to confined quark matter at density $\rho$ and corresponding 
temperature $T$
can be written assuming ultrarelativistic fermionic 
gas also in the confined phase, 
with the additional component corresponding to confinement 
potential, obtained after combining the equations \ref{uconf} and 
\ref{pconf} 

\begin{equation}
4 N_q ( \frac{\eta(4)}{\eta(3)} + 
\frac{g_g}{g_q} \frac{\zeta(4)}{\eta(3)} ) T_0
= 4 N_q \frac{\eta(4)}{\eta(3)} T
+ 2 N_q \frac{w}{6} \kappa \rho^{-1/3}
\end{equation}

and using the relation \ref{durgs} 
one arrives to equation 

\begin{equation}
4 ( \frac{\eta(4)}{\eta(3)} + 
\frac{g_g}{g_q} \frac{\zeta(4)}{\eta(3)} ) T_0
= 4 \frac{\eta(4)}{\eta(3)} 
\frac{h}{2 (\pi g_q \eta(3))^{1/3}} 
\rho^{1/3}
+ \frac{w}{3} \kappa \rho^{-1/3}
\label{isenthnum}
\end{equation}

which after assuming numerical values 
$T_0 = 170 \hbox{ }\rm{MeV}$ (as suggested in the Refs. \cite{HeinzRev}, 
\cite{RandQGP}), $\kappa = 1000 \hbox{ }\rm{MeV/fm}$ \cite{Lund}, 
$g_q = 24$ and $g_q = 16$ leads to a numerical equation 
which has solutions for $w \leq 1.862$.  
The solutions of Eq. (\ref{isenthnum}) are shown in the panel Fig. \ref{fg8}a 
as a dependence 
of the parameter $w$ on quark density. A low-density branch (solid line) 
and high-density branch (dashed line) can be identified, 
separated by the point at $\rho = 0.921 \hbox{ }\rm{fm^{-3}}$ 
corresponding to maximum value of the parameter $w = 1.862$. 
The resulting dependence of 
energy density on quark density 
is shown in the panel Fig. \ref{fg8}b  
and one can see that the solution 
for $w = 1.862$ corresponds approximately to initial energy density 
of quark-gluon plasma at $T_0 = 170 \hbox{ }\rm{MeV}$. 
The low-density branch corresponds to lower energy densities 
and thus represents possible states to which the system can transform 
since it corresponds to both 
quark and energy densities below the phase transition. 
The high-density branch represents the 
states with higher energy densities than the critical density 
and thus can be considered unphysical since it is not obvious that 
the confinement potential is applicable to such states.
From the Eq. (\ref{isenthnum}) thus implies that the phase transition 
can be in principle continuous in the energy density, but 
discontinuous in the quark density. 
After reaching the critical energy density, further 
expansion will be achieved by transforming an increasing 
amount of the quark-gluon plasma into confined quark matter with 
the maximum of about two thirds of quark density, temperature below 150 MeV 
and negative pressure (see Fig. \ref{fg8}c). 
However, once the strings will fragment and a hadronic gas will be formed, 
potential energy will be released, pressure will 
increase dramatically and one can assume an explosion scenario.  
One can imagine a sudden expansion of the bubble of hadronic 
gas to the radius reaching the dimensions of the whole system. 
Such a scenario can offer a natural explanation to the 
"HBT puzzle" where the extracted radii are essentially 
equal for longitudinal, sideward and outward direction and suggest 
an isotropic source with a short emission time.   
HBT puzzle can thus be interpreted as a signature of the phase 
transition. It is also interesting to note formal similarity 
with the cosmologic inflation, a scenario leading to 
isotropic causality-connected universe, which is supposed 
to start from the patch of matter in the state of false vacuum at 
negative pressure. In the present case the confined quark matter 
could represent the false vacuum while the hadronic gas could be considered as 
the true vacuum. Thus, the ultrarelativistic nucleus-nucleus collisions 
may exhibit, apart from the "Little Bang" scenario \cite{LittleBang} and 
possibility of the existence of mini black holes \cite{MiniHoles}, another 
cosmological analogy in the form of "mini-inflation", as a possible 
explanation of the HBT puzzle.  

It is also interesting to compare the results for the quark matter 
to the liquid-gas phase transition at sub-saturation density. Both 
systems will evolve spatial inhomogeneity with one phase 
at positive pressure and the other one at negative pressure, but different 
asymptotic behavior of the potential, one diverging at small distances 
while the other one at large distances, leads to reversion of some 
properties. While in the liquid-gas phase transition the dense 
liquid phase transforms into dilute gaseous phase, in the 
deconfinement-confinement phase transition the dense gaseous phase 
transforms into dilute liquid phase, which subsequently 
transforms into hadronic gas. The liquid consistence of the confined 
quark matter can be judged e.g. from its possible one-dimensional structure, 
as suggested by existing isoenthalpic solutions.   
In both cases the liquid phase is unstable toward 
subsequent decay by fragmentation mechanism. 
It is also interesting to investigate whether 
introduction of strangeness into the model description will lead to asymmetric 
phases. One can assume that for the strange quarks 
the string tension will increase and thus it will be preferable 
for strange particles to remain in the deconfined phase. The strangeness 
of the deconfined phase will increase and the strange hadrons 
will be formed in the latest stage of hadronization.  

\section*{Summary and conclusions}

Thermodynamical properties of the nuclear matter at sub-saturation densities 
were investigated using a simple van der Waals-like equation of state 
with an additional term representing symmetry energy. 
For the isolated system an enthalpy conservation rule was 
introduced resulting in significant limitations for 
the isospin-asymmetric liquid-gas phase transition, which 
is a sizable effect for very isospin-asymmetric systems 
but becomes negligible for symmetric systems. First-order 
phase transition thus appears restricted to asymmetric systems 
while for symmetric systems and symmetric remnants of the 
first-order phase transition the multifragmentation 
decay can exhibit similarity with second-order phase transition. 
The density dependence of the symmetry 
energy scaling with density dependence of the Fermi energy 
satisfactorily describes the symmetry energy at sub-saturation 
nuclear densities. 

The limitations resulting from the isolated system were investigated 
also for the deconfinement-confinement phase transition from the quark-gluon 
plasma to quark matter with the confinement potential. 
It appears that the transition 
can be continuous in energy density while discontinuous in quark 
density. A transitional state of confined quark matter has 
a negative pressure and explosive scenario 
can take place during hadronization. Resulting scenario 
can offer an explanation for the 
HBT puzzle as a signature of the phase transition. 

This work was supported through grant of Slovak Scientific Grant Agency
VEGA-2/5098/25.


\begin{thebibliography}{00}

\bibitem{MVCorrSig}
M. Veselsky and S.J. Yennello, Nucl. Phys. A 749, 114c (2005).

\bibitem{MVIsoTrnd}
M. Veselsky, Fiz. Elem. Chastits At. Yadra 36, 400 (2005); 
Physics of Part. and Nuclei 36, 213 (2005).

\bibitem{Elliot}
J.B. Elliot et al., Phys. Rev. C 62, 64603 (2000).

\bibitem{Zipf}
Y.G. Ma, Phys. Rev. Lett. 83, 3617 (1999).

\bibitem{NimCrit}
Y.G. Ma et al., Phys. Rev. C 69, 031604 (2004).

\bibitem{HeinzRev}
U. Heinz, eprint arXiv:hep-ph/0407360.

\bibitem{Fluid}
U. Heinz, eprint arXiv:nucl-th/0512051.

\bibitem{MachCone}
J.G. Ulery (STAR Collaboration), Nucl. Phys. A 783, 511 (2007).

\bibitem{HBT}
S.S. Adler et al., Phys. Rev. Lett. 93, 152302 (2004); 
J. Adams et al., Phys. Rev. C 71, 044906 (2005).

\bibitem{QCDDrop}
W.N. Zhang and C.Y. Wong, eprint arXiv:hep-ph/0702120.

\bibitem{Sauer}
G. Sauer, H. Chandra and U. Mosel, Nucl. Phys. A 264, 221 (1976).

\bibitem{Muller}
H. M\"uller and B.D. Serot, Phys. Rev. C 52, 2072 (1995).

\bibitem{JBNLimTmp} 
J.B. Natowitz et al., Phys. Rev. C 65, 34618 (2002).

\bibitem{SpinCont}
J. Margueron, P. Chomaz, Phys. Rev. C 67, 41602(R) (2003).

\bibitem{SMM}
J.P. Bondorf et al., Phys. Rep. 257, 133 (1995).

\bibitem{QCDLatt}
E. Laermann and O. Philipsen, Ann. Rev. Nucl. Part. Sci. 53, 163 (2003).

\bibitem{RandQGP}
J. Randrup, Phys. Rev. Lett. 92, 122301 (2004).

\bibitem{Lund}
B. Andersson et al., Phys. Rep. 97, 31 (1983).

\bibitem{LittleBang} 
U. Heinz, Nucl. Phys. A 685, 414c (2001).

\bibitem{MiniHoles}
H. Satz, eprint arXiv:hep-ph/0612151.


\end{thebibliography}
\end{document}